\documentclass[twocolumn,aps,pra,reprint,groupedaddress]{revtex4}
\usepackage{amsmath,amssymb,amsthm,graphicx,hyperref}
\usepackage[FIGTOPCAP,raggedright,nooneline,bf,footnotesize]{subfigure}
\usepackage{indentfirst}
\usepackage{braket}
\usepackage{xcolor}
\usepackage{diagbox}
\usepackage{amssymb}
\usepackage{bbold}

\newtheorem{thm}{Theorem}

\usepackage{hyperref}
\hypersetup{
    colorlinks=true,
    linkcolor=blue,
    filecolor=magenta,      
    urlcolor=cyan,
}

\renewcommand{\eqref}[1]{Eq.~\ref{#1}}

\newcommand{\tr}{{\mathrm{Tr}}}

\begin{document}

\title{Quantum state tomography for qutrits subject to laser cooling}
\author{Artur Czerwinski}
\email{aczerwin@umk.pl}
\affiliation{Institute of Physics, Faculty of Physics, Astronomy and Informatics \\ Nicolaus Copernicus University, Grudziadzka 5, 87-100 Torun, Poland} 

\date{\today}

\begin{abstract}
In this article we propose a dynamic quantum state tomography model for qutrits subject to laser cooling. We prove that one can reduce the number of distinct measurement setups required for state reconstruction by employing the stroboscopic approach. The results are in line with current advances in quantum tomography where there is a strong tendency to investigate the optimal criteria for state reconstruction. We believe that the stroboscopic approach can be considered an efficient tool for density matrix identification since it allows to determine the minimal number of distinct observables needed for quantum state tomography.
\end{abstract}

\maketitle

\section{Introduction}

The term \textit{quantum tomography} is used in reference to a wide variety of methods which aim to reconstruct the accurate representation of a quantum system by performing a series of measurements. Mathematically, the complete knowledge about the state of a quantum system can be encoded in, for example, the density operator, the wavefunction or the Wigner function. In this article we discuss the problem of the density matrix reconstruction. 

One of the most fundamental approaches to quantum state tomography, the so-called static tomography model, enables to reconstruct the density matrix of a quantum system provided one can measure $N^2 - 1$ distinct observables (where $N = dim \mathcal{H}$). Any density matrix can be decomposed in the basis of $SU(N)$ generators in such a way that the coefficients correspond to the mean values of the operators \cite{genki03}. This approach has been excessively studied in many papers and books, such as \cite{altepeter04,alicki87}. However, there is a significant disadvantage connected with this method. In a laboratory one usually is not able to define $N^2 - 1$ distinct physical quantities that could be measured.

The most important property that all tomography models should possess is practicability, which means that a theoretical model should have a potential to be implemented in an experiment in the future. Therefore, when dealing with quantum state tomography we should bear in mind the limitations related to laboratory reality. For this reason, in this article we employ the stroboscopic approach to quantum tomography, which for the first time was proposed by Andrzej Jamiolkowski in \cite{jam83}. Later it was developed in other research papers such as \cite{jam00} and \cite{jam04}. In order to get a broad perspective one may also refer to a very well-written review paper\cite{jam12}. Recently some new results concerning the stroboscopic approach has been presented in \cite{czerwin16a,czerwin17}.

The stroboscopic tomography concentrates on determining the optimal criteria for quantum tomography of open systems. The main goal of this method is to reduce the number of distinct observables required for quantum tomography by utilizing knowledge about time evolution of the system. The data for the density matrix reconstruction is provided by mean values of some hermitian operators $\{ Q_1, \dots, Q_r\}$, where naturally  $Q_i = Q_i^*$. The set of observables is not informationally complete, which means that a single measurement of each operator does not provide sufficient information for quantum state reconstruction.

The underlying principle behind the stroboscopic approach claims that if one has the knowledge about the evolution of the system, each observable can be measured repeatedly at a certain number of time instants. Naturally, each individual measurement is performed over a distinct copy of the system since we do not consider the collapse of the quantum state caused by measurements. Therefore, we assume that our source can prepare a large sample of systems in the identical (but unknown) quantum state. 

In the stroboscopic approach to quantum tomography the fundamental question that we are interested in concerns the minimal number of distinct observables required for quantum state reconstruction. One can recall the theorem concerning the minimal number of observables \cite{jam00}.

\begin{thm}\label{cyclicity}
For a quantum system with dynamics given by a master equation of the form \cite{gorini76,lindblad76}:
\begin{equation}\label{eq:kossakowski}
\dot{\rho} (t) = \mathbb{L} [\rho(t)]
\end{equation}
one can calculate the minimal number of distinct observables for quantum tomography from the formula:
\begin{equation}
\eta := \max \limits_{\lambda \in \sigma (\mathbb{L})} \{ dim Ker (\mathbb{L} - \lambda \mathbb{I})\},
\end{equation}
where by $\sigma (\mathbb{L})$ one should understand the spectrum of the operator $\mathbb{L}$.
\end{thm}

The linear operator $\mathbb{L}$ that appears in the equation \eqref{eq:kossakowski} shall be called the generator of evolution. The number $\eta$ is usually referred to as \textit{the index of cyclicity} of a quantum system.

The theorem \ref{cyclicity} means that for any linear generator $\mathbb{L}$ there exists a set of observables $\{Q_1, \dots, Q_{\eta}\}$ such that their expectation values determine the initial density matrix. Consequently, they also determine the complete trajectory of the state (one can compute the density matrix at any time instant).

If we denote the number of required measurements of each observable from the set $\{ Q_1, ..., Q_{\eta} \}$ by $M_i$ for $i=1,\dots, \eta$, then one can also recall the theorem on the upper limit of moments of measurement \cite{jam04}.

\begin{thm}\label{instants}
In order to provide sufficient data for the density matrix reconstruction the number of times that each observable from the set $\{ Q_1, ..., Q_{\eta} \}$ should be measured satisfies the inequality:
\begin{equation}
 M_i \leq \deg \mu(\mathbb{L}),
\end{equation}
where by $\mu(\mathbb{L})$ we denote the minimal polynomial of $\mathbb{L}$.
\end{thm}

The theorem \ref{instants} gives the upper boundary concerning the number of measurements of each single observable. One can notice that the ability to compute the minimal polynomial of the generator $\mathbb{L}$ is crucial in order to determine the upper limit for the number of measurements. Naturally, another problem relates to the choice of the time instants. Some considerations about this issue can be found in \cite{jam04}.

In the next section the theorems concerning the stroboscopic tomography shall be applied to three level quantum systems with the evolution known as laser cooling. This article brings substantial advancement to the field of quantum state tomography. In \cite{czerwin16a} the author introduced optimal criteria for quantum tomography of qubits. In the current work we proceed towards higher dimensional Hilbert space. We prove that the stroboscopic tomography can be an effective method of state reconstruction for qutrits provided one knows how the system evolves.

\section{Quantum tomography schemes for three level systems subject to laser cooling}

\subsection{Static approach to quantum tomography of qutrits}

In case of three level quantum systems one would naturally employ the Gell-Mann matrices in order to decompose any density matrix. We follow the original notation from \cite{gellmann62} and, therefore, the Gell-Mann matrices shall be denoted by $\{\lambda_1,\lambda_2, \dots, \lambda_8\}$. They have the following forms:

$$ \lambda_1 = \left [ \begin{matrix} 0 & 1  & 0 \\ 1 & 0 & 0 \\ 0 & 0 &0 \end{matrix} \right], \hspace{0.2cm} \lambda_2 = \left [ \begin{matrix} 0 & -i  & 0 \\ i & 0 & 0 \\ 0 & 0 &0 \end{matrix} \right], \hspace{0.2cm}\lambda_3 = \left [ \begin{matrix} 1 & 0 & 0 \\ 0 & -1 & 0 \\ 0 & 0 &0 \end{matrix} \right], $$
$$ \lambda_4 = \left [ \begin{matrix} 0 & 0  & 1 \\ 0 & 0 & 0 \\ 1 & 0 &0 \end{matrix} \right], \hspace{0.25cm} \lambda_5 = \left [ \begin{matrix} 0 & 0  & -i \\ 0 & 0 & 0 \\ i & 0 &0 \end{matrix} \right], \hspace{0.25cm} \lambda_6 = \left [ \begin{matrix} 0 & 0 & 0 \\ 0 & 0 & 1 \\ 0 & 1 &0 \end{matrix} \right], $$
$$ \lambda_7 = \left [ \begin{matrix} 0 & 0 & 0 \\ 0 & 0 & -i \\ 0 & i &0 \end{matrix} \right], \hspace{0.35cm}  \lambda_8 =  \frac{1}{\sqrt{3}}\left [ \begin{matrix} 1 & 0 & 0 \\ 0 & 1 & 0 \\ 0 & 0 & -2 \end{matrix} \right].$$

The Gell-Mann matrices are the generators of the $SU(3)$ group. They are the generalization of the Pauli operators for three level systems. They have some algebraic properties which are useful for quantum state tomography, i.e.:
\begin{equation}
\lambda_i = \lambda_i ^*, \hspace{0.6cm} Tr \lambda_i =0 \hspace{0.3cm} \text{and} \hspace{0.3cm} Tr \lambda_i \lambda_j = 2 \delta_{\ij}.
\end{equation}

For three level quantum systems the initial density matrix $\rho(0) \in \mathcal{S(H)}$ can be decomposed in the basis of the Gell-Mann matrices \cite{genki03}:
\begin{equation}\label{e3.4}
\rho(0) = \frac{1}{3} \mathbb{I}_3 + \frac{1}{2} \sum_{i=1}^8 \langle \lambda_i \rangle \lambda_i,
\end{equation}
where $\langle \lambda_i \rangle$ is the expectation value of the observable $\lambda_i$. Mathematically, it be computed as $\langle \lambda_i \rangle = Tr\{\lambda_i \rho(0)\}$.

If one would like to directly apply the decomposition \eqref{e3.4} in order to reconstruct the density matrix, one would have to know the mean values of $8$ distinct observables: $\{\lambda_1,\lambda_2, \dots, \lambda_8\}$. Such data would be necessary to complete the formula for $\rho(0)$. This approach to quantum tomography, which does not take advantage of the knowledge about evolution, shall be referred to as the static approach. This scheme appears impractical since one is not able to define $8$ distinct physical quantities. This observation justifies the need for more economic approach which aims to decrease the number of distinct observables.

\subsection{Dynamic approach to quantum tomography of qutrits}

Laser cooling is a very widely investigated topic in modern Physics, e.g. \cite{bartana93,tannor99}. A lot of attention has been paid to different aspects of this problem. In this paper we search for a link between laser cooling and quantum state tomography.

An example often studied in the area of laser spectroscopy is a quantum system subject to laser cooling with three energy levels ($dim \mathcal{H} = 3$) \cite{sklarz04}. The evolution of the density matrix of such a three level system is given by a master equation of the form:
\begin{equation}\label{eq:1}
\begin{aligned}
\frac{ d \rho(t)}{d t} = {}& - i [H(t), \rho(t)] + \gamma_1 \left ( E_1 \rho (t) E_1 ^* - \frac{1}{2} \{ E_1  ^* E_1, \rho (t) \} \right )\\&+ \gamma_2 \left ( E_2 \rho (t) E_2^* - \frac{1}{2} \{ E_2^* E_2 , \rho (t) \} \right ),
\end{aligned}
\end{equation}
where $ E_1 = \ket{1} \bra{2} $ and $ E_2 = \ket{ 3} \bra{2}$. The vectors $\{ \ket{1}, \ket{2}, \ket{3}\}$ denote the standard basis in $\mathcal{H}$.

This kind of dynamics appears when the excited state $\ket{2}$ decays spontaneously into two ground states $\ket{1}$ and $\ket{3}$ with corresponding decoherence rates: $\gamma_1$ and $\gamma_3$.

Moreover in this analysis we take $ H(t) = [0] $, where $[0]$ denotes a $3-$dimensional matrix with all entries equal $0$. This assumption means that we shall analyze only the Lindbladian part of the evolution equation.

In case of a three level open quantum system with dynamics given by the master equation from \eqref{eq:1} we can formulate and prove a theorem which provides the minimal number of distinct observables required for quantum tomography.

\begin{thm}\label{cyclicityindex}
For a quantum system subject to laser cooling according to \eqref{eq:1} there exist four distinct observables such that their average values (measured at selected time instants over different copies of the system) suffice to determine the initial density matrix $\rho(0)$.
\end{thm}

\begin{proof}
Based on the method of matrix vectorization \cite{vector,czerwin16a} the dissipative part of the generator of evolution \eqref{eq:1} can be explicitly expressed as a matrix:
\begin{equation}\label{eq:3}
\begin{aligned}
\mathbb{L} = {}& \gamma_1 \left ( E_1 \otimes E_1 - \frac{1}{2} ( \mathbb{I}_9 \otimes E_1 ^T E_1 + E_1 ^T E_1 \otimes \mathbb{I}_9) \right ) +\\ &+  \gamma_2 \left ( E_2 \otimes E_2 -\frac{1}{2} ( \mathbb{I}_9 \otimes E_2 ^T E_2 + E_2^T E_2 \otimes \mathbb{I}_9 ) \right ).
\end{aligned}
\end{equation}

Taking into account the fact that the vectors $\{\ket{1}, \ket{2}, \ket{3}\}$ constitute the standard basis, the matrix form of the quantum generator $\mathbb{L}$ can be obtained:
\begin{equation}\label{eq:4}
\mathbb{L} = \left[ \begin{matrix} 0 & 0 & 0 & 0 & \gamma_1& 0 & 0 & 0 & 0  \\
 0 & -\Gamma & 0 & 0 & 0 & 0 & 0 & 0 & 0  \\
 0 & 0 & 0 & 0 & 0 & 0 & 0 & 0 & 0  \\
 0 & 0 & 0 &- \Gamma& 0 & 0 & 0 & 0 & 0 \\
0 & 0 & 0 & 0 & - 2\Gamma & 0 & 0 & 0 & 0 \\
 0 & 0 & 0 &0 & 0 & -\Gamma & 0 & 0 & 0  \\
 0 & 0 & 0 & 0 & 0 & 0 & 0 & 0 & 0  \\
 0 & 0 & 0 &0 &0 & 0 & 0 & -\Gamma  & 0  \\
0 & 0 & 0 & 0 & \gamma_2 & 0 & 0 & 0 & 0  \end{matrix} \right ],
\end{equation}
where $ \Gamma = \frac{1}{2} (\gamma_1 + \gamma_2 )$.

Having the matrix form of the generator of evolution $ \mathbb{L}$, one can calculate its eigenvalues:
\begin{equation}\label{eq:5}
\begin{aligned}
\sigma ( \mathbb{L}) =  \left\{ 0, 0, 0,  0, -2 \Gamma , -\Gamma, -\Gamma, -\Gamma, -\Gamma \right \}.
\end{aligned}
\end{equation}

Since in this case the operator $ \mathbb{L}$ is not self-adjoint, the algebraic multiplicity of an eigenvalue does not have to be equal to its geometric multiplicity. But one can quickly determine that there are four linearly independent eigenvectors that correspond to the eigenvalue $0$. Therefore, we can find the index of cyclicity for the operator in question:
\begin{equation}\label{eq:6}
\eta = \max \limits_{\lambda \in \sigma (\mathbb{L})} \{ dim Ker (\mathbb{L} - \lambda \mathbb{I}_9) \} = 4,
\end{equation}
which means that we need exactly four distinct observables to perform quantum tomography of the analyzed system.
\end{proof}

One can instantly notice that if the static approach was applied to three level laser cooling, one would have to measure $8$ distinct observables whereas in the dynamic approach $4$ observables suffice to perform quantum tomography. If one thinks of potential applications in experiments, then our result means that one would have to prepare $4$ different experimental setups instead of $8$. This observation demonstrates that the stroboscopic approach has an advantage over the static approach because it is more economic when it comes to the number of distinct kinds of measurement.

The next issue that we are interested in is the minimal polynomial for the operator $ \mathbb{L}$. Assuming that this polynomial has the monic form, i.e.:
\begin{equation}\label{eq:7}
d_3 \mathbb{L} ^ 3 + d_2 \mathbb{L} ^ 2+d_1 \mathbb{L} + d_0 \mathbb{I} = 0,
\end{equation}
one can get :
\begin{equation}\label{eq:8}
d_3 = 1,\hspace{0.1cm} d_2 = \frac{3}{2} (\gamma_1 + \gamma_2), \hspace{0.1cm} d_1 = \frac{1}{2} (\gamma_1 + \gamma_2)^2,\hspace{0.1cm} d_0 = 0.
\end{equation}

Thus, we see that $\deg \mu(\mathbb{L})  = 3$. This means that each observable should be measured at most at three different time instants. One can conclude that, since we need $8$ independent pieces of information to reconstruct the initial density matrix, not every observable will be measured the maximum number of times. To provide a precise answer to the question concerning the algebraic structure of the observables and the choice of time instants we shall accept additional assumptions concerning the generator of evolution.

Let us consider a special case of the generator of evolution defined in \eqref{eq:4} such that $\gamma_1 = \frac{1}{4} $ and $\gamma_2 = \frac{3}{4}$. For this specific generator we can formulate a theorem.
\begin{thm}\label{observables}
The initial density matrix $\rho(0)$ of a three level system subject to laser cooling can be reconstructed from the mean values of four observables of the form:
\begin{equation}
\begin{split}
& Q_1 = \left [ \begin{matrix} 1 & 0 & 0 \\ 0 & -1 & 1+i \\ 0 & 1-i &0 \end{matrix} \right], \hspace{0.2cm} Q_2 = \left [ \begin{matrix} 0 & 0 & 1+i  \\ 0 & 0 & 0 \\ 1-i & 0 & 0 \end{matrix} \right],\\
& Q_3 = \left [ \begin{matrix} 0 & 1 & 0 \\ 1 & \frac{1}{\sqrt{3}} & 0 \\ 0 & 0 & - \frac{2}{\sqrt{3}} \end{matrix} \right], \hspace{0.4cm} Q_4 = \left [ \begin{matrix} 0 & i & 0  \\ -i & 0 & 0 \\ 0 & 0 & 0 \end{matrix} \right],\\
\end{split}
\end{equation}
where the mean values of $Q_1$ and $Q_2$ are measured at $3$ distinct time instants and the observables $Q_3$ and $Q_4$ once at $t=0$.
\end{thm}

\begin{proof}
According to the assumptions of the stroboscopic tomography the information that one can obtain from an experiment is encoded in the mean values of some observables, which mathematically can be written as:
\begin{equation}
m_i (t_j) = \tr \{ Q_i \rho(t_j) \},
\end{equation}
where $\rho(t_j) = \exp (\mathbb{L} t_j ) [\rho(0)]$.

One is aware that $\exp (\mathbb{L} t_j )$ can be decomposed as:
\begin{equation}
\exp (\mathbb{L} t ) = \alpha_0 (t) \mathbb{I}_9 + \alpha_1 (t) \mathbb{L} + \alpha_2 (t) \mathbb{L}^2,
\end{equation}
where the functions $\{\alpha_0 (t), \alpha_1 (t), \alpha_2 (t) \}$ are linearly independent. In order to determine these functions we need to employ the minimal polynomial of $\mathbb{L}$ and then solve a system of differential equations \cite{jam04,czerwin16a}. Having done the necessary computations, one gets:
\begin{equation}
\begin{cases}
\alpha_0 (t) = 1\\
\alpha_1 (t) = e^{-t} - 4 e^{-\frac{1}{2} t} + 3\\
\alpha_2 (t) = 2 e^{-t} - 4 e^{-\frac{1}{2} t} +2
\end{cases}
\end{equation}

Since one is able to decompose $\exp (\mathbb{L} t_j )$ in the basis of three operators $\{\mathbb{I}_9, \mathbb{L} , \mathbb{L}^2,\}$ due to linearity of the matrix trace we get:
\begin{equation}
\begin{aligned}
m_i (t_j) = {}& \alpha_0 (t_j) \tr \{ Q_i \rho(0) \} + \alpha_1 (t_j) \tr \{ Q_i \mathbb{L}[\rho(0)] \}+\\& + \alpha_2 (t_j) \tr \{ Q_i \mathbb{L}^2[\rho(0)] \}.
\end{aligned}
\end{equation}

If by $\mathbb{L}^*$ we shall denote the dual operator to $\mathbb{L}$, then by changing the perspective from the Schrödinger picture to the Heisenberg representation we can obtain:
\begin{equation}
\begin{aligned}
m_i (t_j) = {}& \alpha_0 (t_j) \tr \{ Q_i \rho(0) \} + \alpha_1 (t_j) \tr \{ \mathbb{L}^*[Q_i] \rho(0) \}+\\& + \alpha_2 (t_j) \tr \{ (\mathbb{L}^*)^2[Q_i] \rho(0) \}.
\end{aligned}
\end{equation}

This means that if the mean value of the observable $Q_1$ is measured at three distinct time instants one gets a matrix equation:
\begin{equation}\label{matrixq1}
\left[ \begin{matrix} m_1 (t_1) \\ m_1 (t_2) \\ m_1 (t_3) \end{matrix} \right] = 
\left[\begin{matrix} \alpha_0 (t_1) & \alpha_1 (t_1)  & \alpha_2 (t_1)  \\  \alpha_0 (t_2) & \alpha_1 (t_2)  & \alpha_2 (t_2) \\  \alpha_0 (t_3) & \alpha_1 (t_3)  & \alpha_2 (t_3) 
\end{matrix} \right] \left[ \begin{matrix} \tr \{ Q_1 \rho(0)\} \\ \tr \{ \mathbb{L}^*[Q_1] \rho(0)\} \\ \tr\{ (\mathbb{L}^*)^2[Q_1] \rho(0)\} \end{matrix} \right].
\end{equation}

One can observe that since the functions $\{\alpha_0 (t), \alpha_1 (t), \alpha_2 (t) \}$ are linearly independent if one selects three different non-zero time instants such that $t_1 \neq t_2 \neq t_3$, then the matrix $[\alpha_k (t_j)]$ must be invertible. It implies that the measurement results $\{ m_1 (t_1), m_1 (t_2) , m_1 (t_3)\}$ can be translated into a set of scalar products: $\{  \tr \{ Q_1 \rho(0)\}, \tr \{ \mathbb{L}^*[Q_1] \rho(0)\}, \tr\{ (\mathbb{L}^*)^2[Q_1] \rho(0)\}$.

The very same measurement procedure, which must result in a matrix equation analogous to \eqref{matrixq1}, can be performed for the observable $Q_2$. Triple measurement of $Q_2$ at distinct time instants yields a set of the following scalar products: $\{ \tr \{ Q_2 \rho(0)\}, \tr \{ \mathbb{L}^*[Q_2] \rho(0)\}, \tr\{ (\mathbb{L}^*)^2[Q_2] \rho(0)\}$.

Finally, a single measurement of the average value of $Q_3$ and $Q_4$ at time instant $t=0$ provides another two scalar products: $\{ \tr \{Q_3\rho(0)\} , \tr \{ Q_4 \rho(0)\}$.

One can check numerically that the operators $\{\mathbb{I}_3, Q_1, \mathbb{L}^*[Q_1], (\mathbb{L}^*)^2[Q_1], Q_2, \mathbb{L}^*[Q_2], (\mathbb{L}^*)^2[Q_2],  Q_3,  Q_4 \}$ constitute a spanning set (they are linearly independent), which means that:
\begin{equation}\label{crucial}
\begin{aligned}
\textit{Span}{}&\{\mathbb{I}_3, Q_1, \mathbb{L}^*[Q_1], (\mathbb{L}^*)^2[Q_1], Q_2, \mathbb{L}^*[Q_2], (\mathbb{L}^*)^2[Q_2],  Q_3,  Q_4 \}\\&= B_*(\mathcal{H}).
\end{aligned}
\end{equation}

The equation \eqref{crucial} expresses the necessary and sufficient condition for the ability to reconstruct the initial density matrix of a qutrit subject to laser cooling. This condition is satisfied for the observables defined in the theorem \ref{observables}, which can be observed numerically by using the software Mathematica 11.

In other words, the operators: $$\{\mathbb{I}_3, Q_1, \mathbb{L}^*[Q_1], (\mathbb{L}^*)^2[Q_1], Q_2, \mathbb{L}^*[Q_2], (\mathbb{L}^*)^2[Q_2],  Q_3,  Q_4 \}$$ constitute a quorum, i.e. they span the space to which $\rho(0)$ belongs. Therefore, the scalar products that one can calculate from the measurement results can be considered a complete set of information. Thus, the measurement procedure, which utilizes only $4$ distinct kinds of measurement, provides $8$ independent pieces of information which are sufficient for the density matrix reconstruction.
\end{proof}

The theorems \ref{cyclicityindex} and \ref{observables} provide a complete description of the quantum tomography scheme. One knows exactly what steps should be taken in order to compute the unknown density matrix. The results are in accord with current trends in quantum state tomography where a lot of attention is paid to the methods which aim to reduce the experimental effort, e.g. \cite{czerwin16b,oren17}. If one can access the knowledge about dynamics of the system encoded in the generator of evolution, it seems more convenient to perform repeatedly the same kind of measurement (over distinct copies of the system) rather than develop a large number of different experimental setups.

\section{Summary}
In this paper we presented a complete quantum tomography model for qutrits subject to laser cooling. The stroboscopic approach was applied to determine the optimal criteria for density matrix reconstruction. It was demonstrated that one can reduce the number of distinct observables by $50\%$ provided the knowledge about evolution is applied. The algebraic structure of the observables was presented along with a detailed description of the scheme. Dynamic methods of state reconstruction appear to be very practical since they allow to retrieve the initial density matrix in the most economic way, by minimizing the number of distinct measurement setups.

The current work can be extended in the future research by studying the problem of quantum state tomography for systems subject to laser cooling with more than three energy levels.

\end{document}